\documentclass{aa}  
\usepackage{graphicx}
\usepackage{txfonts}
\usepackage{natbib,pjournal}

\begin{document}
\title{The Cosmic Far-Infrared Background Buildup Since Redshift 2 at
  70 and 160 microns in the COSMOS and GOODS fields.}

   \titlerunning{The CIB buildup since z=2 at 70 and 160~$\mu$m in COSMOS and GOODS.}

   \author{M. Jauzac\inst{1,2,3}
          \and
          H. Dole\inst{2,3}
          \and
          E. Le Floc'h\inst{4}
          \and
          H. Aussel\inst{4}
          \and
          K. Caputi\inst{5}
         \and
          O. Ilbert\inst{1}
          \and
          M. Salvato\inst{6,7}
          \and
          N. Bavouzet\inst{2,3}
          \and
          A. Beelen\inst{2,3}
          \and
          M. B\'ethermin\inst{2,3}
          \and
          J.-P. Kneib\inst{1}
          \and
          G. Lagache\inst{2,3}
          \and
          J.-L. Puget\inst{2,3}
          }

\authorrunning{M. Jauzac et al.}

   \offprints{M. Jauzac \email{mathilde.jauzac@oamp.fr}}

   \institute{Laboratoire d'Astrophysique de Marseille, Universit\'e de Provence, CNRS, 13388 Marseille Cedex 13 France
     \and
     Univ Paris Sud, Institut d'Astrophysique Spatiale (UMR 8617), b\^at 121, Orsay, F-91405, France 	
     \and
     {CNRS, Orsay, F-91405, France} 
     \and
     IRFU, SAp, CNRS, Saclay, b\^at 709, Orme des merisiers, 91191 Gif-sur-Yvette France
    \and
     SUPA, Institute for Astronomy, The University of Edinburgh, Royal Observatory, Edinburgh - EH9 3HJ, UK
     \and
     California Institute of Technology, MC 105-24, 1200 East California Boulevard, Pasadena, CA 91125
      \and
      Max-Planck Institute for Plasma Physics \& Cluster of Excellence, Boltzmann Strasse 2, Garching 85748 Germany
   }

   \date{Received ; accepted}

  \abstract
  {The Cosmic Far-Infrared Background (CIB) at wavelengths around $
    160~\mu$m corresponds to the peak intensity of the whole
    Extragalactic Background Light, which is being measured with
    increasing accuracy. However, the build up of the CIB emission as
    a function of redshift, is still not well known.}
  {Our goal is to measure the CIB history at 70~$\mu$m and 160~$\mu$m
    at different redshifts, and provide constraints for infrared galaxy
    evolution models.}
  {We use complete deep Spitzer 24~$\mu$m catalogs down to about
    80~$\mu$Jy, with spectroscopic and photometric redshifts
    identifications, from the GOODS and COSMOS deep infrared surveys
    covering 2 square degrees total.  After cleaning the Spitzer/MIPS
    70~$\mu$m and 160~$\mu$m maps from detected sources, we stacked
    the far-IR images at the positions of the 24~$\mu$m sources
    in different redshift bins.  We measured the contribution of each
    stacked source to the total 70 and 160~$\mu$m light, and compare
    with model predictions and recent far-IR measurements made with
    Herschel/PACS on smaller fields. }
  {We have detected components of the 70 and 160~$\mu$m backgrounds in
    different redshift bins up to $z \sim 2$.  The contribution to the
    CIB is maximum at $0.3 \le z \le 0.9$ at 160$\mu$m (and $z \le
    0.5$ at 70~$\mu$m).  A total of 81\% (74\%) of the 70 (160)~$\mu$m
    background was emitted at $z<1$. We estimate that the AGN relative
    contribution to the far-IR CIB is less than about 10\% at $z<1.5$.
    We provide a comprehensive view of the CIB buildup at 24, 70, 100,
    160~$\mu$m.}
  {IR galaxy models predicting a major contribution to the CIB at
    $z<1$ are in agreement with our measurements
   .  The consistency of our results with those
    obtained through the direct study of Herschel far-IR data at 160~$\mu$m confirms that the stacking analysis method is a valid
    approach to estimate the components of the far-IR background using
    prior information on resolved mid-IR sources.  Our results are
    available online http://www.ias.u-psud.fr/irgalaxies/ .  [(...) abstract abridged]}


   \maketitle
%
\section{Introduction}

The Extragalactic Background Light (EBL) is the relic emission of
galaxy formation and evolution, i.e. due to star formation and
accretion processes (with this definition, the Cosmic Microwave
Background due to recombination at redshift $z\sim$ 1100 is not part
of the EBL). The EBL spectrum peaks in the Far-Infrared (FIR), where
it is commonly known as the Cosmic Infrared Background (CIB)
\cite[]{puget96,hauser98,hauser2001,kashlinsky2005,dole2006}.  The EBL
and the CIB encode the emission processes of structure formation, and
can thus be used to constrain the photon budget of the cooling
processes leading the baryons to fall within the dark matter
halos and form galaxies. The measurements of the EBL level and
structure bring thus one of the many useful constraints for the
models.

The CIB Spectral Energy Distribution is known with increasing accuracy
\cite[for instance in the FIR and submillimetre regime:
]{puget96,aharonian2006,dole2006,bethermin2010}, but little is known
about its history, i.e. its buildup as a function of redshift.  This
missing information should help constrain galaxy evolution models, and
also better understand the physics of blazars, whose high-energy
photons interact with the CIB along the line of sight
\cite[e.g. ]{aharonian2007,albert2008,raue2009,kneiske2009}.

The history of the CIB buildup can be derived by integrating the
luminosity functions of galaxies as a function of redshift (neglecting
other sources of diffuse emission and thus assuming that the CIB is
due to galaxies). This is a very difficult task in practice, since
high-redshift luminosity functions are not yet measured at wavelengths
close to the peak of the CIB (near 160~$\mu$m) but instead in the
mid-infrared range \cite[e.g. ]{lefloch2005,caputi2007}, or only in
the local universe \cite[]{soifer91,takeuchi2006}. This situation is
about to change with the latest Spitzer surveys and the ongoing deeper
Herschel surveys \cite[]{magnelli2009,clements2010,dye2010}.

\begin{figure}[!t]
\centering \includegraphics[width=0.49\textwidth]{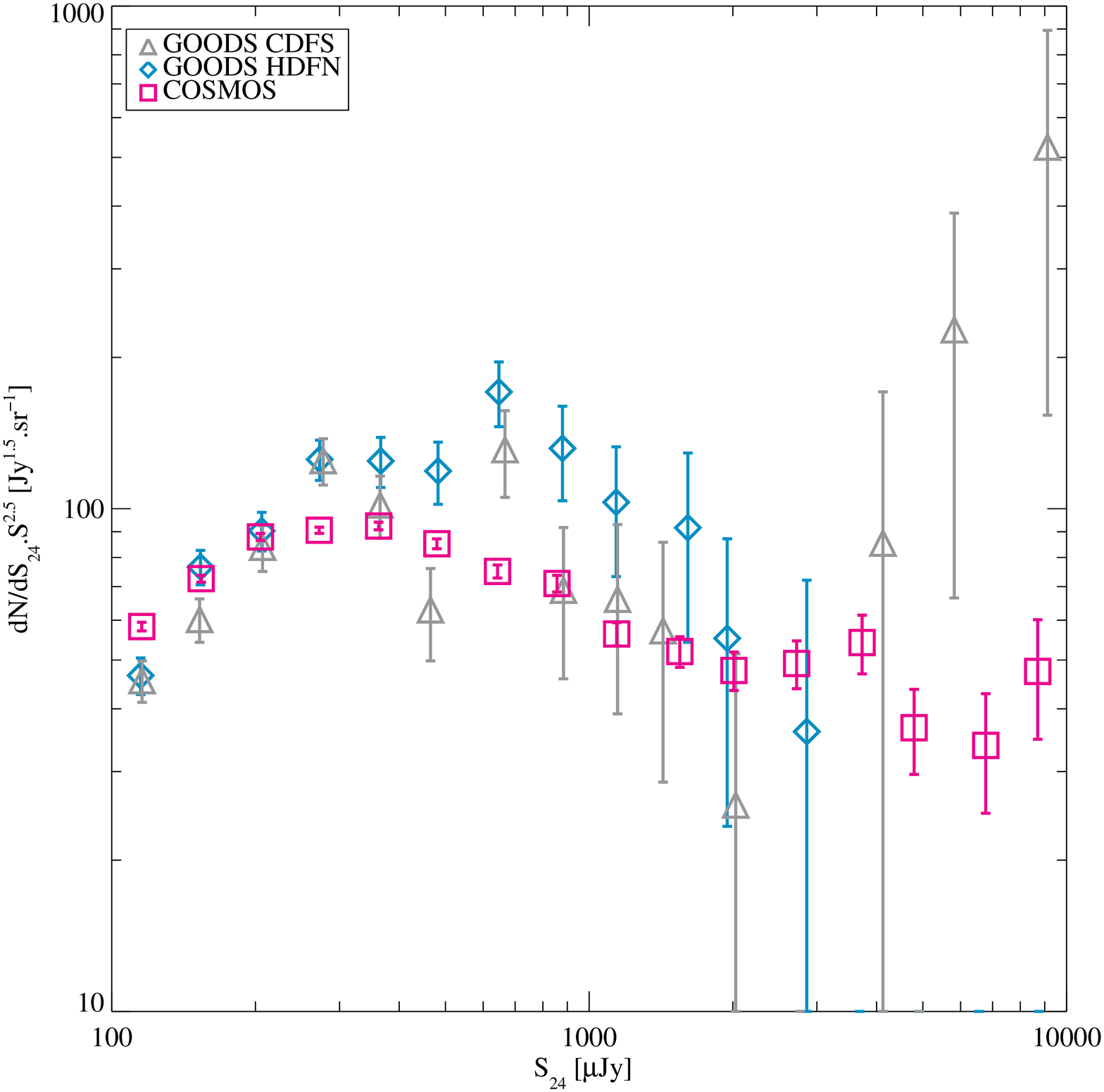}
\caption{Number counts at 24~$\mu$m in the GOODS HDFN (blue
    diamond), GOODS CDFS (gray triangle) and COSMOS (pink square)
    fields. The errors bars used only include Poisson statistics, and
    not cosmic variance.} \label{fig:counts}
\end{figure}

Two recent breakthroughs have been made by using COSMOS and GOODS
surveys.  Firstly, using about 30000 Spitzer 24~$\mu$m selected
sources with accurate photometric redshifts \cite[]{ilbert2009},
\cite{lefloch2009} were able to measure the 24~$\mu$m background
buildup with redshift (e.g. their figures 7 to 9). They furthermore
show that the redshift information is crucial when confronting data to
the models, since it helps breaking degeneracies. Secondly, using the
redshift identification of Herschel/PACS 100 and 160~$\mu$m sources,
\cite{berta2010} were able to measure the CIB build up in four
redshift bins, in the 140~arcmin$^2$ GOODS-N field (an area about 40
times smaller than used in this analysis).

In this paper, we measure the 70~$\mu$m and 160~$\mu$m CIB history
since $z=2$, with the use of a stacking analysis of galaxies detected
at 24~$\mu$m (a good proxy for the 160~$\mu$m CIB population, e.g.
\cite{dole2006,bethermin2010}) in the Spitzer data of the GOODS and
COSMOS fields. This approach complements on a large area what is being
done with Herschel in \cite{berta2010} at 100 and 160~$\mu$m.

%
\begin{figure}[!t]
\centering \includegraphics[width=0.45\textwidth]{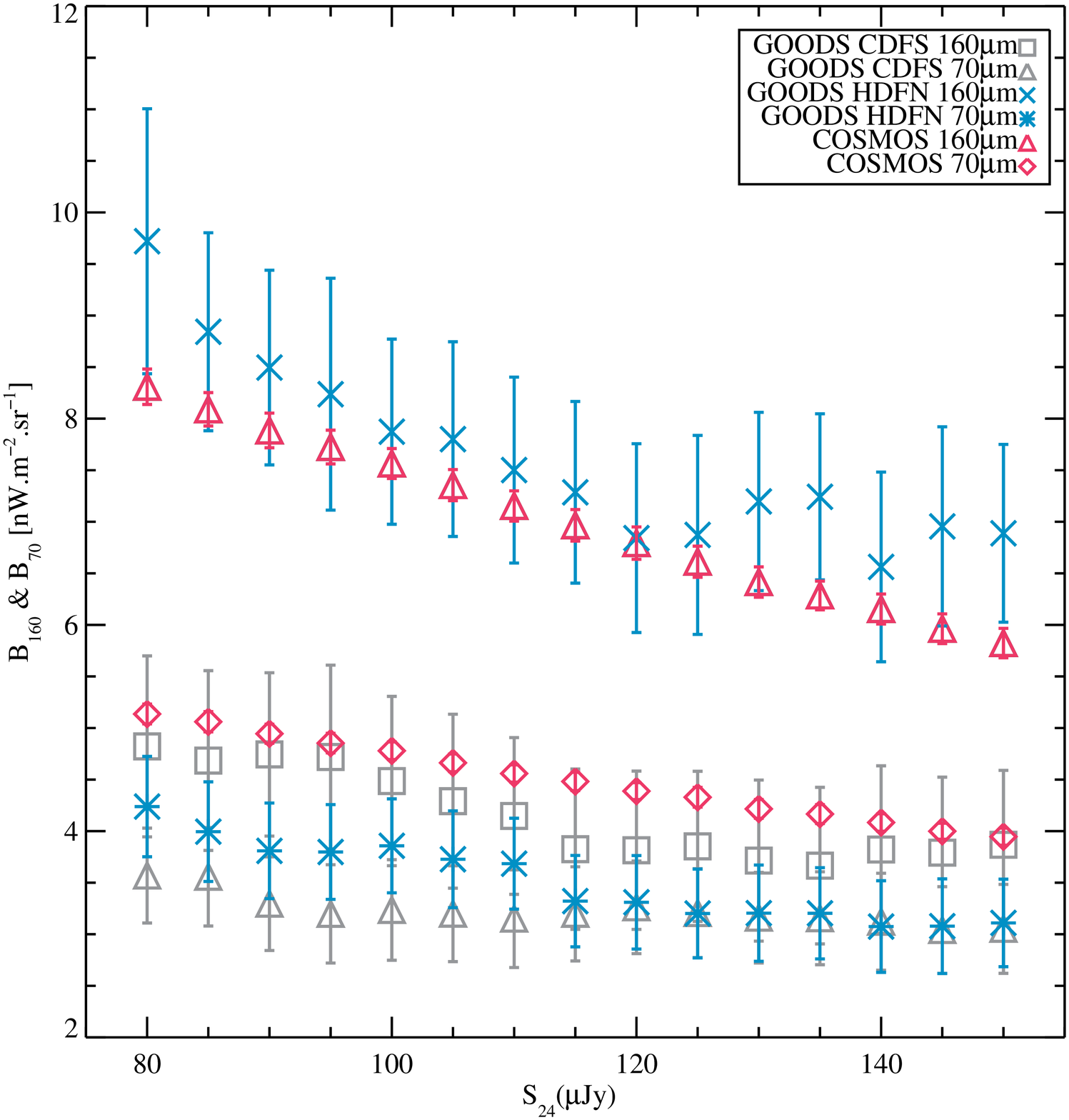}
\caption{ Cumulative stacked brightness at 70~$\mu$m and
    160~$\mu$m (in nW.m$^{-2}$.sr$^{-1}$) on the CLEANed maps, as a
    function of the 24~$\mu$m flux of our sample, regardless of the
    redshift. the 3 fields are represented at 160 and 70~$\mu$m:
    COSMOS; GOODS-N (HDFN) and GOODS-S (CDFS).}
\label{fig:stacked_flux_vs_s24}
\end{figure}

%
\begin{figure*}[!ht]
\centering \includegraphics[height=0.20\textheight]{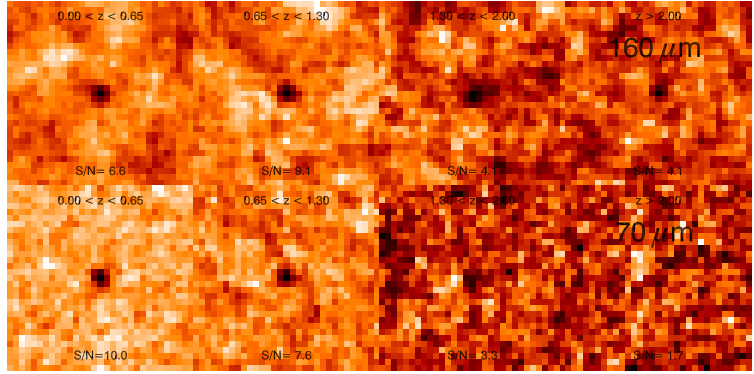}
\caption{Images of all the stacked galaxies on the 160~$\mu$m (top)
  and 70~$\mu$m (bottom) GOODS CLEANed maps by redshift bin (right to
  left): 0$<$z$<$0.65; 0.65$<$z$<$1.3; 1.3$<$z$<$2 and z$>$2. The S/N
  ratio is indicated in each image. Notice the detection in the two
  first redshift bins at both wavelengths. Images are $305 \times 305$
  sq. arcsec. wide at 70~$\mu$m (with 9.85 arcsec pixel plate), and
  $496 \times 496$ sq. arcsec. at 160~$\mu$m (with 16 arcsec pixel
  plate). The PSF FWHM being 18 arcsec. (resp. 40) at 70~$\mu$m
  (resp. 160~$\mu$m), the PSF shown on these figures has about the
  same extend of 1.8 to 2.5 pixels at both wavelengths.  }
\label{fig:stacked_imagesGOODS}
\end{figure*}

\begin{figure*}[!ht]
\centering \includegraphics[width=0.99\textwidth]{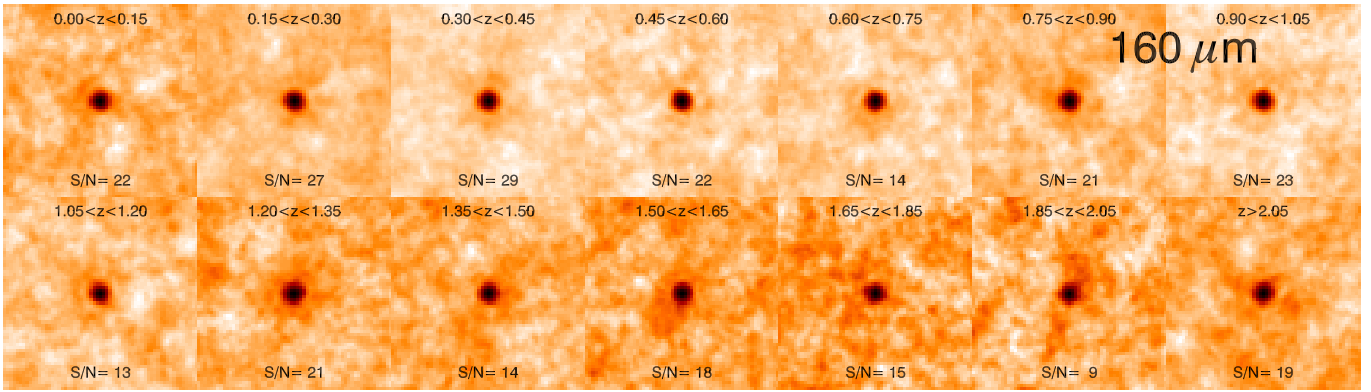}
\caption{Images of all the stacked galaxies on the 160~$\mu$m COSMOS
  CLEANed maps with 14 redshift bins (left to right and top to
  bottom). The S/N ratio is indicated in each image. Notice the
  detection in all redshift bins at both wavelengths. Images are $488
  \times 488$ sq. arcsec. at 160~$\mu$m (with 8 arcsec pixel
  plate). The PSF FWHM of 40 arcsec. corresponds to 5 pixels on these
  images.}
\label{fig:stacked_imagesCOSMOS}
\end{figure*}

\begin{figure*}[!ht]
\centering \includegraphics[width=0.99\textwidth]{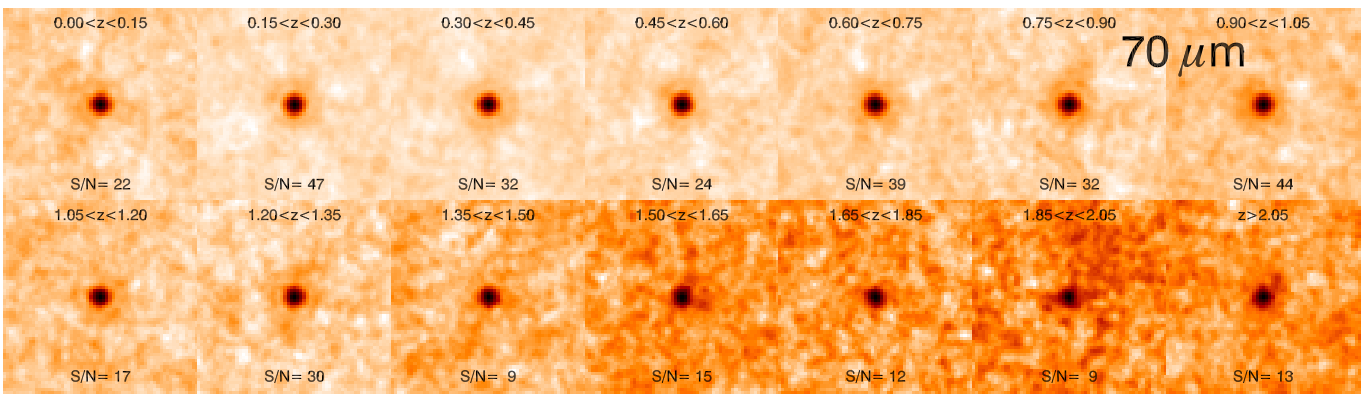}
\caption{Images of all the stacked galaxies on the 70~$\mu$m COSMOS
  CLEANed maps with 14 redshift bins. Images are $244 \times 244$
  sq. arcsec. wide at 70~$\mu$m (with 4 arcsec pixel plate). The PSF
  FWHM of 18 arcsec. corresponds to 4.5 pixel on these images.  }
\label{fig:stacked_imagesCOSMOS70}
\end{figure*}

\section{Data and Sample}
\subsection{GOODS data}
\label{sect:data:goods}
The data were acquired by the MIPS imaging photometer at 24~$\mu$m,
70~$\mu$m and 160~$\mu$m \cite[]{rieke2004} onboard the Spitzer
infrared space telescope \cite[]{werner2004}, and come from the GOODS
team \cite[]{chary2004} and Guaranteed Time observations
\cite[]{papovich2004,dole2004a} of the Chandra Deep Field South (CDFS)
and the Hubble Deep Field North (HDFN).  \cite{papovich2004} extracted
a catalog at 24~$\mu$m, with 80\% completeness at 80~$\mu$Jy.  We use
a sample of 1349 galaxies with 24~$\mu$m flux densities $S_{24} \ge
80~\mu$Jy, located in the two GOODS fields, north and south, for a
total area of 291 Sq. Arcmin \cite[]{caputi2006,caputi2007}. The
galaxies have been completely identified, and redshifts determined for
all of them, with more than 45\% of spectroscopic redshifts. Active
Galactic Nuclei (AGN) are separated from the star-forming systems
using X-ray data and near-infrared (3.6 to 8~$\mu$m) colors: we have
136 AGNs for 1213 star-forming systems \cite[]{caputi2007}.  For our
purpose of measuring the contribution of mid-infrared galaxies to the
far-infrared background by redshift slice, we cut the 24~$\mu$m sample
in four redshift bins: 0$<$z$<$0.65 with 317 sources (of which 9
AGNs), 0.65$<$z$<$1.3 with 575 sources (45 AGNs), 1.3$<$z$<$2 with 259
sources (38 AGNs) and z$>$2 with 198 sources (44 AGNs). These bins
have been chosen to maximize the number of sources present in each
bin, while keeping the same $\Delta z$ width.


\begin{table*}[!t]
  \caption[]{Number of 24~$\mu$m sources per redshift bins for the COSMOS field. $N_{tot}$ is the total number of galaxies used in
    the stacks, and  $N_{AGNs}$ is the number of sources identified as AGN that were used in the stacks to estimate the AGN contribution.}
\label{tab:cosmos_number}
\begin{center}
\begin{tabular}[h]{lcccccccc}
\hline\\[-5pt]
z bin & 0$<$ $z$ $<$0.15 & 0.15$<$z$<$0.3 & 0.3$<$ $z$ $<$0.45 & 0.45$<$z$<$0.6 & 0.6$<$ $z$ $<$0.75 & 0.75$<$z$<$0.9 & 0.9$<$ $z$ $<$1.05  \\
\hline\\[-5pt]
$N_{tot}$ & 2083 & 1559 & 2853 & 2201 & 3225 & 3590 & 3478  \\ 
$N_{AGNs}$  & 34   & 32   & 74   & 48   & 88   & 110  & 123  \\

\hline\\[-5pt]

z bin  & 1.05$<$z$<$1.2 & 1.2$<$ $z$ $<$ 1.35 & 1.35$<$z$<$1.5 & 1.5$<$z$<$1.65 & 1.65$<$z$<$1.85 & 1.85$<$z$<$2.05 & $z$ $>$2.05 \\
\hline\\[-5pt]
$N_{tot}$ & 2670 & 1401 & 2044 & 1311 & 1519 & 2073 & 2833 \\ 
$N_{AGNs}$  & 83   & 76   & 55   & 225  & 232  & 200  & 288 \\
\hline\\[-5pt]
\end{tabular}
\end{center}
\end{table*}

\subsection{COSMOS data}
\label{sect:data:cosmos}
The Cosmic Evolution survey (COSMOS) data were acquired by MIPS at 24,
70 and 160~$\mu$m.  The 24~$\mu$m observations of the COSMOS field is
part of two General Observer programs (PI D. Sanders): G02 (PID 20070)
carried out in January 2006, G03 (PID 30143) carried out in 2007. We
use a total net area of 1.93 Sq. degrees.  \cite{lefloch2009}
extracted a catalogue at 24~$\mu$m and provided us a sample of 32840
galaxies with 24~$\mu$m flux densities $S_{24} \ge 80~\mu$Jy. The
  completeness limit is at the order of 90\% at this level. (Notice
  that the survey sensitivities are similar in the COSMOS and GOODS
  fields, at 24, 70 and 160~$\mu$m). The 24~$\mu$m galaxies have been
completely identified, and redshifts derived by \cite{ilbert2009} and
\cite{salvato2009} for the optically and X-Ray selected sources of the
COSMOS field. We use the \cite{salvato2009} photometric redshift
catalogue of the \cite{cappelluti2009} X-Ray sources catalogue,
optically matched by \cite{brusa2007} \cite[]{brusa2010} to identify
the AGN in the COSMOS field (Le Floc'h et al., 2009). Notice that the
X-ray flux limits used in the soft (0.5-2keV), hard (2-10keV) or
ultra-hard (5-10keV) bands are $5\times10^{-16}$, $3\times10^{-15}$
and $5\times10^{-15}$ erg cm$^{-2}$ s$^{-1}$, respectively
\cite[]{cappelluti2007,cappelluti2009,salvato2009}.  We complete this
sample with sources with a power-law SED \cite[]{alonso-herrero2006}
in the redshift range 1.5$<$z$<$2.5 using IRAC colors (at lower and
higher redshifts, the colors can be contaminated by the PAH or stellar
bumps) in the same way as for the GOODS sample.  We obtained 1668
sources (1115 X-Ray sources, 553 power-law sources) detected at
24~$\mu$m and identified as AGNs for 31172 star forming systems.

The COSMOS sample being larger than the GOODS one we used 14
redshift bins, described in table~\ref{tab:cosmos_number}.  The source
statistics in these fields is summarized in Fig.~\ref{fig:counts},
showing the number counts of the GOODS survey (CDFS and HDFN) as well
as the COSMOS field, corrected for incompleteness. The errors bars
used only include Poisson statistics, and not cosmic variance, and are
thus likely underestimated. There is no evidence of a relative major
over- or under-density, except maybe a slight over-density in the HDFN
around 1 mJy, which has a negligible contribution to the total
background.

\begin{figure*}[!t]
\centering \includegraphics[width=0.99\textwidth]{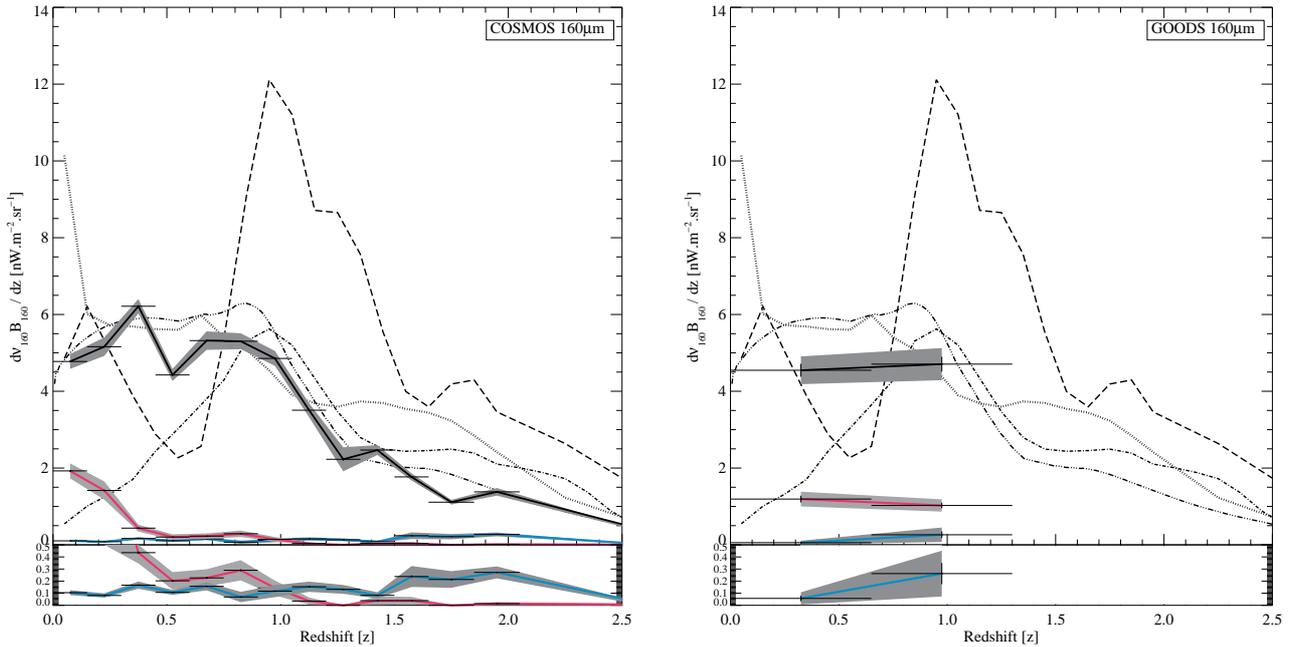}
\caption{Differential 160~$\mu$m background brightness $d\nu _{160}B_
  {160}/dz$ as a function of redshift, in units of
  nW.m$^{-2}$.sr$^{-1}$, in the COSMOS field (left) and GOODS field
  (right). The lower part of the plot shows a linear zoom
    between 0 and 0.5 nW.m$^{-2}$.sr$^{-1}$ in order to exhibit the
    trend of cleaned sources and AGNs. Black solid lines: total
  contribution of infrared galaxies. Solid red lines: contribution
  from resolved sources only. Solid blue line: contribution from AGN
  only. Dash line \cite{lagache2004} model, with $S_{24}>80~\mu$Jy
  cut. Dot line: \cite{le_borgne2009} model, with $S_{24}>80~\mu$Jy
  cut. Dot-dash line:\cite{valiante2009}. Dot-dot-dot-dash
  line: \cite{bethermin2010c} model, with $S_{24}>80~\mu$Jy cut. Models are discussed in Sect.~\ref{sect:models} and figure~\ref{fig:dbdzmodel}.} \label{fig:dbdz160}
\end{figure*}

%
\section{Analysis}
\label{sect:analysis}

\subsection{Stacking analysis}
In order to estimate the contribution of mid-infrared galaxies to the
70~$\mu$m and 160~$\mu$m background, we make use of a stacking
analysis\footnote{The IAS stacking library, written in IDL, is
  publicly available at http://www.ias.u-psud.fr/irgalaxies , cf
  \cite{bavouzet2008a} and \cite{bethermin2010}}
\cite[]{dole2006,bethermin2010b}. This method consists in stacking the
70 and 160~$\mu$m maps at the location of the galaxies detected at
24~$\mu$m. The use of this method is justified by two reasons. 1) The
24~$\mu$m population is a good proxy for the 70~$\mu$m and 160~$\mu$m
populations making up most of the CIB near its peak
\cite[]{dole2006,bethermin2010}. 2) Only a few sources are
individually detected at FIR wavelengths and do not resolve much of the
background
\cite[]{dole2004a,frayer2006,frayer2006a,dole2006,frayer2009}.  Notice
that stacking may suffer from the contamination of galaxy clustering,
since the stacked image shows in two dimensions the 2 point angular
correlation function
\cite[]{dole2006,bavouzet2008a,bethermin2010}. However with the Spitzer and
Herschel beams, the effects of clustering in the stacking are not
important (less than 15\%)
\cite[]{bavouzet2008a,fernandez-conde2008,fernandez-conde2010}.

We first stack the 70~$\mu$m and the 160~$\mu$m MIPS data
  (CLEANed maps) as a function of the 24~$\mu$m flux, regardless the
redshift of the sources (Fig.~\ref{fig:stacked_flux_vs_s24}). This
allows us to check the consistency of the procedure, since the total
brightness measured for stacks down to $S_{24}=80~\mu$Jy should be
equal to the sum of the brightnesses obtained by redshift slices, as
well as identify possible biases.  The stacks in the COSMOS
  and the two GOODS fields at 70 and 160~$\mu$m show strong
  dependencies on the fields: while COSMOS and GOODS-N (HDFN) stacks
  are consistent within 20\%, stacks in CDFS field is systematically
  lower than COSMOS by a factor of about 1.4-1.8.  The better
statistics in the COSMOS field (surface area and number of sources) is
limiting the impact of the variance due to the large scale structure,
and we attribute the systematic lower values in GOODS-S to this
effect.

Prior to stacking the 24~$\mu$m catalog in redshift bins on the
70~$\mu$m and 160~$\mu$m maps, we use the {\sc clean} algorithm
\cite[]{hogbom74} to subtract the few resolved sources present in the
Far-Infrared maps, in order to remove any bias in the resulting
photometry of the stacked images. The stacking analysis
  presented on Fig.~\ref{fig:stacked_flux_vs_s24} was also done on the
  cleaned Far-Infrared maps. In the COSMOS field, we clean the
sources brighter than 80 mJy (resp. 20 mJy) at 160~$\mu$m
(resp. 70~$\mu$m), level corresponding to 90 to 95\% completeness,
levels computed by Monte-Carlo simulations on the data themselves
\cite[]{bethermin2010}. In both GOODS fields, we removed all
  the detected sources at 160~$\mu$m $\&$ 70~$\mu$m identified at
  24~$\mu$m. It corresponds to sources brighter than 19 mJy at
  160~$\mu$m (5 sources in GOODS HDFN $\&$ 12 sources in GOODS CDFS)
  and 4.4 mJy at 70~$\mu$m (8 sources in GOODS HDFN $\&$ 17 sources in
  GOODS CDFS). These brightest detected sources at 70 and 160~$\mu$m
have been individually identified at 24~$\mu$m without ambiguity, and
the redshift of the 24~$\mu$m source is used. The flux densities of
the removed detected sources are converted into brightnesses, and
added at the very end of the process to account for their CIB
contribution (even if it's a small fraction at far-IR wavelengths).

We estimate the AGNs contribution to the CIB as a function of redshift
using the identifications described in sections~\ref{sect:data:goods}
and \ref{sect:data:cosmos}.

The stacking procedure is performed for each redshift bin
independently, and the images of the stacks are presented in
Fig.~\ref{fig:stacked_imagesGOODS} and
Fig.~\ref{fig:stacked_imagesCOSMOS}, for the GOODS and COSMOS fields
respectively, together with the measured signal-to-noise ratios.

We summarize our approach:
\begin{itemize}
\item compute the brightness (by redshift bin) of the detected sources
  that are removed from the maps to create the cleaned maps;
\item select galaxies at 24~$\mu$m (all of them, or just AGN, or just
  non-AGN), by redshift bin;
\item stack at the positions of the selected galaxies in the 70 and
  160~$\mu$m cleaned maps;
\item perform photometry and bootstrap on those stacks;
\item compute uncertainty budget.
\end{itemize}

All the measurements discussed in this section, i.e. number of stacked
sources, resolved sources, AGN, and resulting brightnesses as a
function of redshift, are summarized in tables
~\ref{tab:CIBcontrib160GOODS}, ~\ref{tab:CIBcontrib070GOODS},
~\ref{tab:CIBcontrib160COSMOS}, and \ref{tab:CIBcontrib070COSMOS}.

\begin{figure*}[!t]
\centering \includegraphics[width=0.99\textwidth]{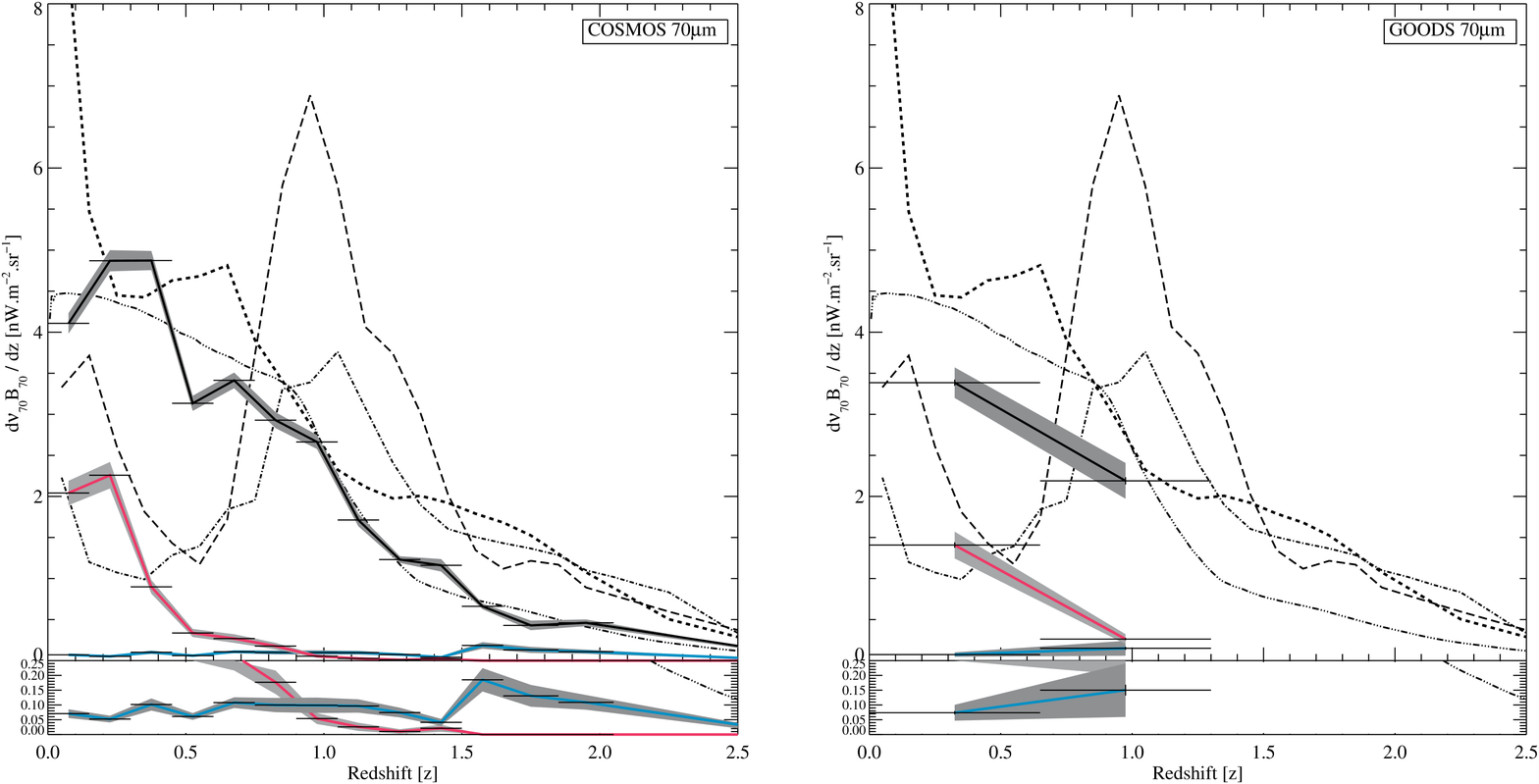}
\caption{Differential 70~$\mu$m background brightness $d \nu_{70}B_
  {70}/dz$ as a function of redshift, in units of
  nW.m$^{-2}$.sr$^{-1}$. in the COSMOS field (left) and GOODS field
  (right). The lower part of the plot shows a linear zoom between
    0 and 0.25 nW.m$^{-2}$.sr$^{-1}$ in order to exhibit the trend of
    cleaned sources and AGNs. Black solid lines: total contribution
  of infrared galaxies. Solid red lines: contribution from resolved
  sources only. Solid blue line: contribution from AGN only. Dash line
  \cite{lagache2004} model, with $S_{24}>80~\mu$Jy cut. Dot line:
  \cite{le_borgne2009} model, with $S_{24}>80~\mu$Jy
  cut. Dot-dash line:\cite{valiante2009}. Dot-dot-dot-dash
  line: \cite{bethermin2010c} model, with $S_{24}>80~\mu$Jy cut. Models are discussed in Sect.~\ref{sect:models} and figure~\ref{fig:dbdzmodel}.}
 \label{fig:dbdz70}
\end{figure*}

\subsection{Photometry and uncertainty estimations}
We perform aperture photometry on the stacked images with the
following parameters at 160~$\mu$m: aperture radius of 25", a sky
annulus to estimate the background between 80 and 110", and an
aperture correction of 2.29. At 70~$\mu$m, the parameters are: 18", 50
and 70", and 1.68.  We have secure detections in all redshift bins at
160~$\mu$m and 70~$\mu$m, except in the two highest redshift bin
($z>1.3$) in GOODS. The signal to noise ratio is better in COSMOS than
in GOODS, due to the larger number of sources used (the number of
sources used are reported in tables~\ref{tab:CIBcontrib160GOODS} to
\ref{tab:CIBcontrib070COSMOS}). We thus will take into account in our
analysis the redshifts bins $0<z<1.3$ in GOODS and $0<z<2.5$ in
COSMOS.

The error bars come from three quadratically summed terms. 1- the
photometry uncertainty. 2- the Poisson noise coming from the number of
stacked sources. 3- a bootstrap analysis. 

The bootstrap analysis is done by running the stacking process $N_{b}$
times (usually $N_{b} = 5000$ and $N_{b} = 14000$ for GOODS and COSMOS
resp.) of a new sample composed of randomly selected sources from our
original sample, keeping the total number of sources constant
\cite[]{bavouzet2008a}; this means that some stacked positions might
be present zero, or multiple times in each realization. The bootstrap
error bar comes from the standard deviation of the distribution of the
photometry measured on these $N_{b}$ realizations. Notice that the
signal-to-noise ratio of the detections in the stacked images (only
photometric) is higher than the value quoted in this paper, since we
add the Poisson and bootstrap terms to estimate the final error bar,
which takes into account the dispersion of the underlying sample.  The
final error bar is thus larger than just the photometric noise
estimate. The error bars on the AGNs samples were determined using a
smaller number of bootstrap, $N_{b} = 100$ and $N_{b} = 2000$ for
GOODS and COSMOS respectively.

The variance due to the large scale structure (also known as cosmic
variance) and field-to-field variations is a systematic component of
the noise, that is difficult to estimate at this stage. The Poisson
noise, used here, gives a strict lower limit of the cosmic variance.

\subsection{Measurements}

By adding the brightness obtained from the stacking of 24~$\mu$m
sources with $S_{24} \ge 80~\mu$Jy and the few detected far-infrared
sources, we measure $B_{160_{tot-GOODS}} = 7.53 \pm 0.52$
nW.$m^{-2}$.$sr^{-1}$ at 160 $\mu$m, $B_{70_{tot-GOODS}} = 3.97 \pm
0.17$ nW.$m^{-2}$.$sr^{-1}$ at 70 $\mu$m and $B_{160_{tot-COSMOS}} =
7.88 \pm 0.19$ nW.$m^{-2}$.$sr^{-1}$ at 160 $\mu$m,
$B_{70_{tot-COSMOS}} = 4.95 \pm 0.08$ nW.$m^{-2}$.$sr^{-1}$ at 70
$\mu$m (see also the summary in Tab.~\ref{tab:CIBtot}).

If we compare with the models from \cite{lagache2004},
\cite{le_borgne2009}, and \cite{bethermin2010c} (cf
section~\ref{sect:models} and Tab.~\ref{tab:CIBtot}) applying the same
selection of using the 24~$\mu$m sources with $S_{24} \ge 80~\mu$Jy,
we obtain that we resolve 66 to 89\% of the 160~$\mu$m background, and
75 to 98\% of the 70~$\mu$m background. 

To be more specific, using the only post-Herschel model in hand
\cite[]{bethermin2010c}, our data show that we resolve in COSMOS 90\%
at 160~$\mu$m and 98\% at 70~$\mu$m of the background we are supposed
to measure with the selection at 24~$\mu$m applied. Our selection
introduces an incompleteness in the CIB estimate due to the fainter
24~$\mu$m sources ($S_{24} < 80~\mu$Jy), missing in our analysis; this
loss implies that we resolve 68\% of the total 160~$\mu$m background
and 81\% of the total 70~$\mu$m background in COSMOS (see
sect.~\ref{sect:models} for the details).  For comparison,
\cite{berta2010} identified about 50\% of the 100 and 160~$\mu$m
backgrounds in individual sources, and account for 50 to 75\% of the
background when stacking at the positions of 24~$\mu$m galaxies, as we
do.

%
\section{Discussion}
In the following, we present the measurements and the models in the
form of $\frac{d(\nu B_{\nu})}{dz}$ versus redshift $z$, where $\nu
B_{\nu}$ is the CIB brightness in nW.m$^{-2}$.sr$^{-1}$, $\lambda$ is
the wavelength (70~$\mu$m or 160~$\mu$m) and $\nu$ the corresponding
frequency. This representation has the advantage of being independent
of the redshift binning, thus allowing a direct comparison between
datasets and models differently sampled in redshift. We discuss data
and models with the prior selection of $S_{24} > 80~\mu$Jy, and will
show (sect.~\ref{sect:models}) that our conclusions for $z<1.5$,
i.e. where most of the FIR background arises, are not modified by this
prior selection compared to taking fainter galaxies.

\subsection{The 160~$\mu$m background: history since $z=2$}

The distribution of the $160~\mu$m CIB measured brightness as a
function of redshift (Fig.~\ref{fig:dbdz160}) shows a plateau between
redshifts 0.3 and 0.9 in COSMOS and GOODS fields, followed by a
decrease at higher redshift.  The small dip at $z=0.5$ in COSMOS is
not significant, since it disappears when increasing the size of the
redshift bin ($\Delta z=0.3$ instead of 0.15) and is likely due to a
structure in the COSMOS field. The GOODS field has the same trend in
redshift.

The contribution from resolved sources is maximum at $z<0.3$ and
strongly decreases at higher redshift, in agreement with the
identifications of \cite{frayer2006}. The AGN contribution is rather
constant with redshift; the relative contribution of AGN thus rises
with redshift. Assuming the COSMOS field is representative of the
whole CIB population, we derive that 33\% of the 160~$\mu$m background
is accounted at redshifts $0 < z < 0.5$, 41\% for $0.5 < z < 1$, 17\%
for $1 < z < 1.5$, and 9\% for $1.5 < z < 2$.  Our results are
consistent with \cite{berta2010}, who analyzed a deep sample in the
GOODS-N field at 160~$\mu$m with PACS/Herschel. Most of the
far-infrared sources are resolved by Herschel, and the stacks of
24~$\mu$m sources provide slightly more depth. Their peak at $z=1$ is
more pronounced than our analysis.

The \cite{lagache2004},
\cite{le_borgne2009},\cite{valiante2009} and
\cite{bethermin2010c} models are overplotted to our measurements in
Fig.~\ref{fig:dbdz160}, using the same selection of $S_{24} \ge
80~\mu$Jy as applied on the data. Pre-Herschel models predict
different redshift distribution of $d\nu _{160}B_{160}/dz$. The
\cite{lagache2004} models peaks at $z\sim 1$, and does not
fit our data.  Our data are in better qualitative agreement with the
\cite{le_borgne2009} model, (except for $z < 0.3$), but none models
fit the $z>1$ tail. The problem of the discrepancy between our data
and the \cite{le_borgne2009} model at $z < 0.3$ might be twofold: our
data lacks of statistics at very low redshift due to the relative
small sky area, and the model might be overpredicting low-z galaxies
because of the lack of a cold component in the galaxies SED used. The
\cite{bethermin2010c} fits well the low, intermediate and high
redshift ranges, likely because it is based on the minimization of
recent Spitzer and Herschel data, and already takes into account the
FIR and submm statistical properties of galaxies (see
Sect.~\ref{sect:models}).

\subsection{The 70~$\mu$m background: history since $z=2$}

The distribution of the $70~\mu$m CIB measured brightness as a
function of redshift (Fig.~\ref{fig:dbdz70}) shows a maximum
contribution $z < 0.5$ in COSMOS, consistent with the GOODS
measurements.  The peak contribution at 70~$\mu$m lies at lower
redshift than at 160~$\mu$m, which is expected as a consequence of the
K-correction (the effect of the redshifted shape of the galaxies
spectra). This effect is also seen between 100~$\mu$m and 160~$\mu$m
in the PACS/Herschel data by \cite{berta2010}. The dip at $z \sim 0.5$
is likely a cosmic variance effect, as it is not seen in the
GOODS field, and it disappears when we use broader redshift bins. This
dip does not change our conclusion about the 70~$\mu$m
background emission with redshift.

The contribution from resolved sources is maximum at $0.15 < z < 0.3$
and strongly decreases at higher redshift, in agreement with the
identifications of \cite{frayer2006}. The AGN contributions is rather
constant with redshift. Assuming the COSMOS field is representative of
the whole CIB population, we derive that 43\% of the 70~$\mu$m
background is accounted at redshifts $0 < z < 0.5$, 38\% for $0.5 < z
< 1$, 13\% for $1 < z < 1.5$, and 5\% for $1.5 < z < 2$.

The \cite{lagache2004} model predicts a peak of the CIB at 70~$\mu$m
at around $z=1$, not seen in the data. The \cite{le_borgne2009}
predicts a peak at lower redshift ($z \le 0.5$), with a strong
contribution at $z \sim 0$, not seen either in the data; otherwise,
the decrease at $z>0.5$ has a shape comparable to the data, despite a
larger high redshift tail. The post-Herschel \cite{bethermin2010c}
model nicely follows observed evolution in COSMOS and GOODS
(Fig.~\ref{fig:dbdz70}).

\begin{figure}[!t]
\centering \includegraphics[width=0.49\textwidth]{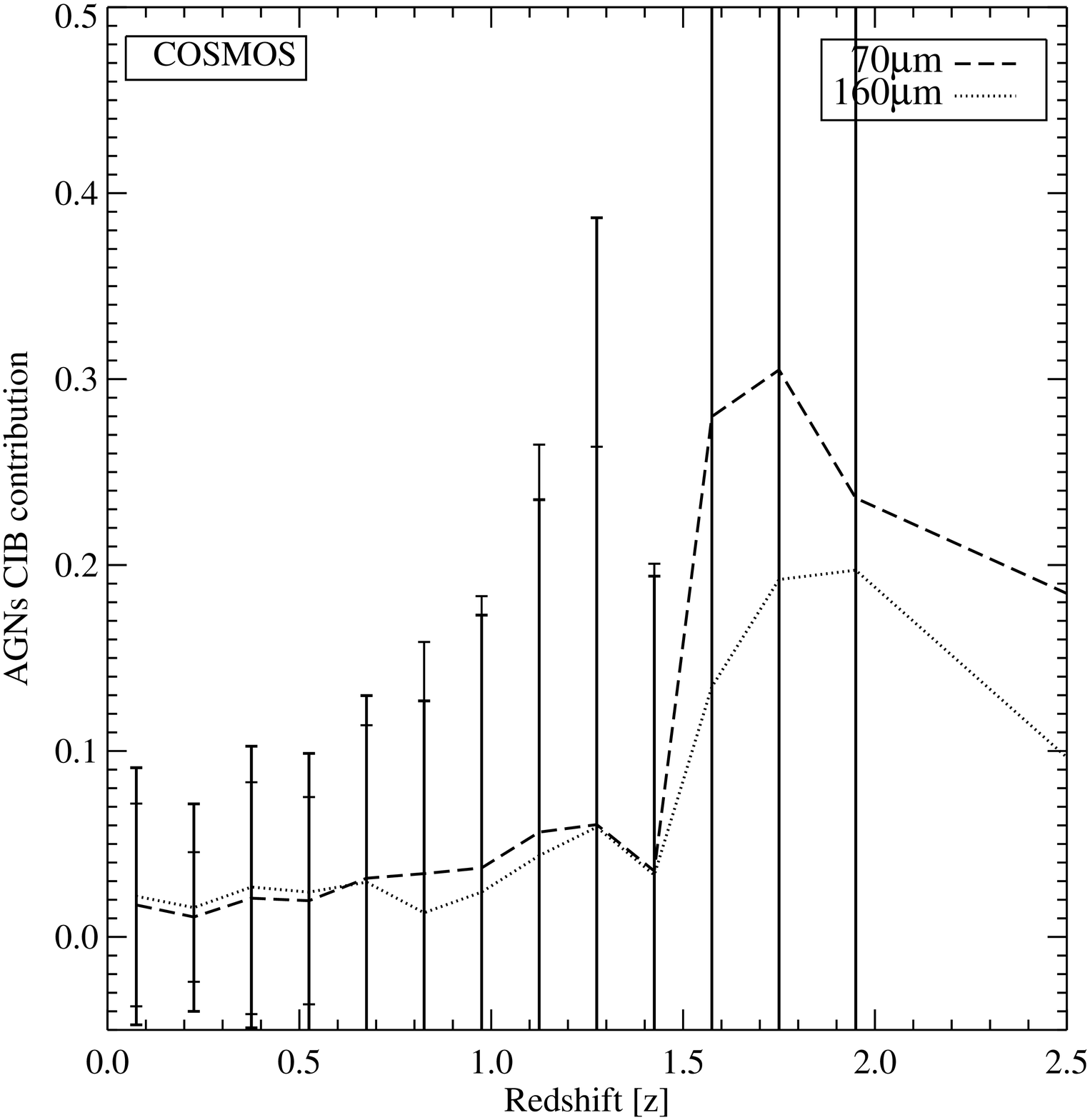}
\caption{The relative contribution of identified AGN to the 70 and
  160~$\mu$m CIB as a function of redshift, with the selection at
  24~$\mu$m with $S_{24} > 80 \mu$Jy.} \label{fig:agncontrib}
\end{figure}

\begin{figure*}[!t]
\centering \includegraphics[width=0.85\textwidth]{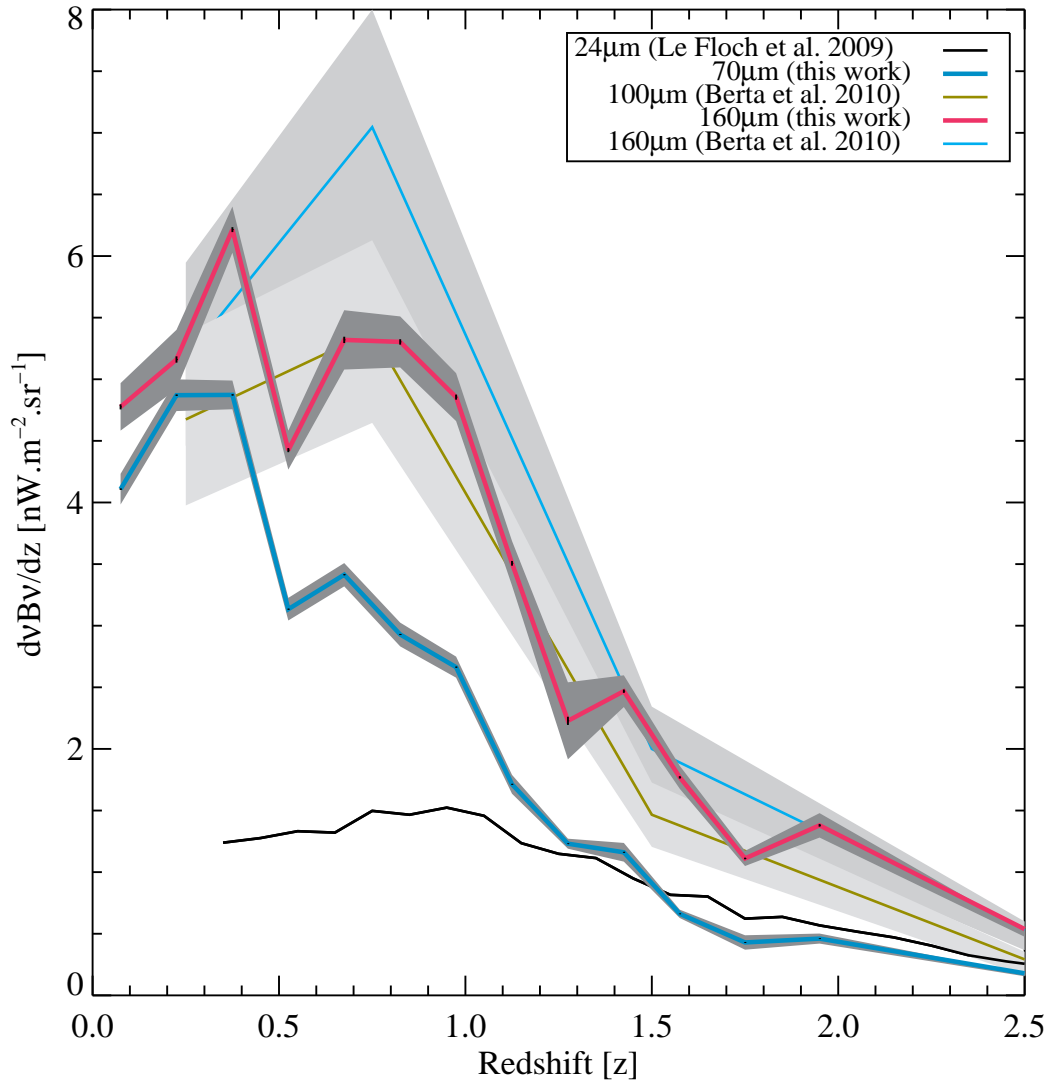}
\caption{The Cosmic Infrared Background history since $z \sim 2$ at
  24, 70, 100 and 160~$\mu$m as measured from galaxies selected at
  24~$\mu$m with $S_{24} > 80 \mu$Jy. 24~$\mu$m data come from
  \cite{lefloch2009}, 70 and 160~$\mu$m from this work, all using the
  Spitzer/MIPS data of COSMOS field.  100 and 160~$\mu$m data in four
  redshift bins come from \cite{berta2010} in GOODS-N with
  Herschel/PACS (and using the same 24~$\mu$m
  prior.).} \label{fig:cib_wavelength}
\end{figure*}

\subsection{Role of AGN}

The AGN contribution in each redshift bin is shown on
Fig.~\ref{fig:dbdz70} and \ref{fig:dbdz160}, as the lower blue area.
The estimate of this contribution can be considered as a ``best
effort'' estimate, because of the difficulty of the task of
identifying the origin of the far-infrared emission (star formation or
AGN). Our AGN identification relies on X-ray detections and IRAC
colors \cite[]{caputi2006,salvato2009}, but the far-infrared emission
is not necessarily physically linked to the AGN \cite[]{lefloch2007}.

Our analysis shows that the absolute contribution of AGN to the CIB at
70 and 160~$\mu$m is rather constant with
redshift. Figure~\ref{fig:agncontrib} shows the relative contribution
to the CIB, and is obtained by dividing the AGN contribution to the
total CIB contribution.  Because of the smaller contribution of higher
redshift sources to the CIB, the AGN fraction contribution to the CIB
is increasing with redshift, from about $3 \pm 10$\% for $0 < z <
1.5$, and up to possibly 15-25\% for $z>1.5$, but our large error bars
 do not allow any meaningful estimate. We thus can only state
that the relative AGN contribution is less than about 10\% at
$z<1.5$. Our results are in agreement with \cite{daddi2007} who
predict that the contribution of AGNs shouldn't exceed more than 7$\%$
up to a redshift unity.

AGN are thought to play a central role in terms of physical processes
driving galaxy evolution and regulating star formation trough feedback
\cite[e.g. ]{magorrian1998,ferrarese2000,bower2006,hopkins2006a,cattaneo2009,hopkins2010}.
However, our work confirms that, in terms of total energy
contributions to the CIB, the AGN play a minor role. This conclusion
is in agreement with the identifications of \cite{frayer2006} and the
analysis of \cite{valiante2009}, where (their figure~19) is shown that
less than 10\% of the sources with $S_{24}> 80~\mu$Jy \cite[i.e. the
sources making-up the CIB at 24 to 160~$\mu$m, see ]{dole2006} have a
significant AGN contribution.

\subsection{The 24, 70, 100 and 160~$\mu$m backgrounds}

The mid- and far-infrared background buildup at 24, 70, 100 and
160~$\mu$m as a function of redshift is summarized in
figure~\ref{fig:cib_wavelength} and is available
online\footnote{http://www.ias.u-psud.fr/irgalaxies/}. The 24~$\mu$m
data come from \cite{lefloch2009} and were normalized to
2.86~nW.m$^{-2}$.sr$^{-1}$ \cite[]{bethermin2010}, and the data at
100~$\mu$m come from \cite{berta2010}, while the data at 70 and
160~$\mu$m come from this work. At wavelengths larger than 60~$\mu$m,
the observed buildup sequence shows an increasing contribution from $z
\sim 0.5$ to $z \sim 1$ with increasing wavelength. This behaviour is
expected, as consequence of the redshifted peak in the galaxy spectral
energy distributions, or k-correction
\cite[e.g. ]{lagache2004,lagache2005}. The 24~$\mu$m buildup
  shows a flatter (or broader) distribution in redshift, with a
  maximum contribution around $z \sim 1$; This mid-infrared
  distribution has a relative $z>1.5$ contribution larger than in the
  far-IR, i.e. the decay slope is smaller at 24~$\mu$m than at
  70~$\mu$m and larger wavelengths.

A detailed comparison, however, is still difficult because of the
cosmic variance. The results from \cite{berta2010} are based on
GOODS-N, an area about 50 times smaller than used here. We furthermore
showed that large scale structure is visible at $z<0.5$ in the COSMOS
field.

The fact that most of the CIB between 100 and 160~$\mu$m is identified
as being produced by $z < 1$ sources is in line with expectations
\cite[e.g. ]{lagache2005,bethermin2010c}. The 24~$\mu$m background
has, however, a significant fraction from galaxies at $z>1$
\cite[30\%, according to ]{lefloch2009}. As expected and observed
\cite[]{marsden2009}, most of the submillimeter background is made of
sources lying at higher redshifts ($z>1.5$). This wavelength versus
redshift dependence of the background can be explained by the sum of
SED of galaxies at various redshifts, in particular the peak emission
in the far-infrared of the reprocessed starlight by dust being
redshifted. Thus, the SED of the CIB is broader and flatter at
far-infrared and submillimeter wavelengths than any individual galaxy
SED.

\begin{figure*}[!t]
\centering \includegraphics[width=0.99\textwidth]{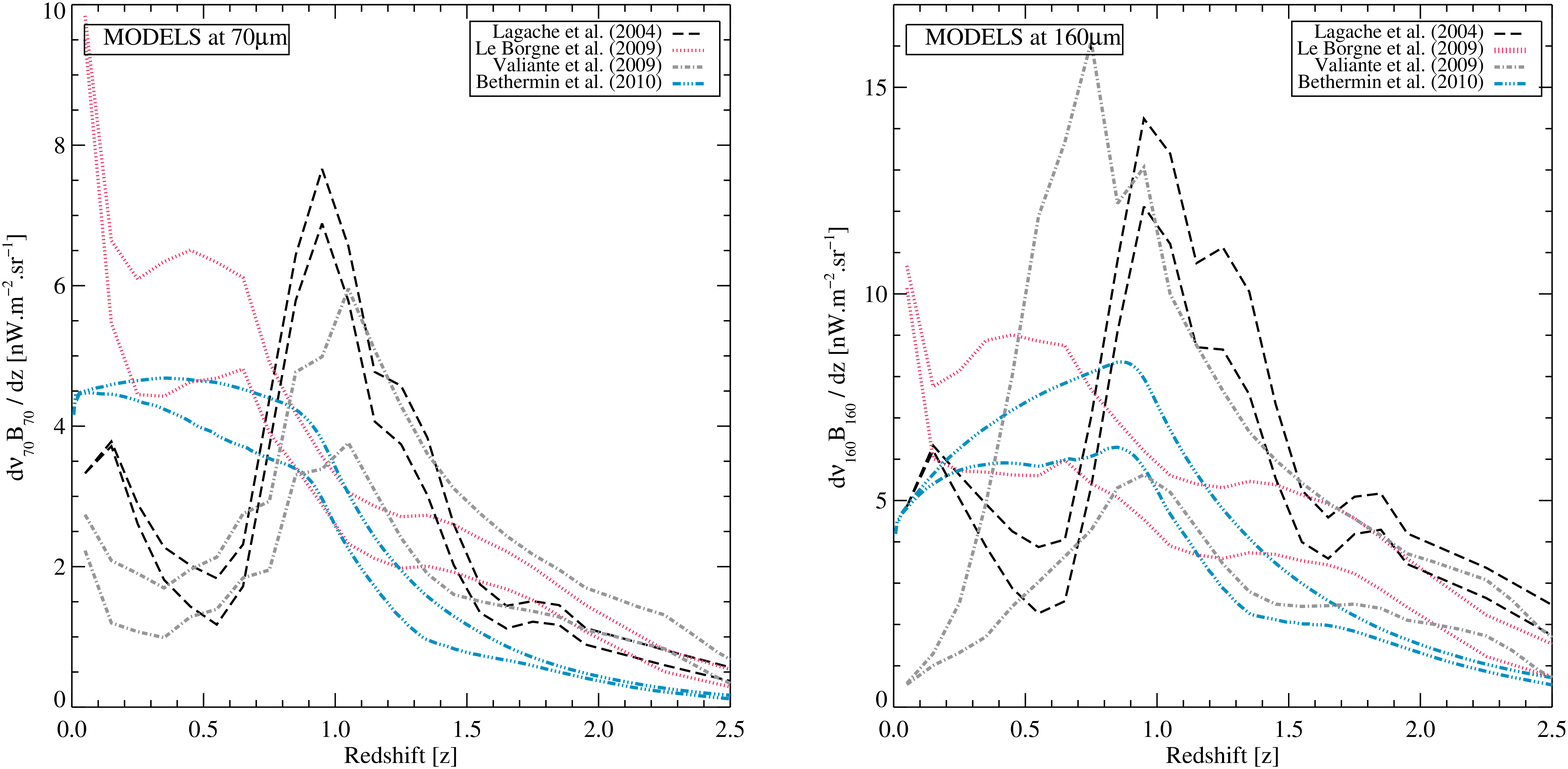}
\caption{Differential 70 (left) and 160~$\mu$m (right) background
  brightness $d\nu _{\nu}B_ {\nu}/dz$ as a function of redshift, in
  units of nW.m$^{-2}$.sr$^{-1}$, as predicted by 4 models: Dash black line:
  \cite{lagache2004} model; Dot magenta line: \cite{le_borgne2009} model; Dot-dash grey line: \cite{valiante2009};
  Dot-dot-dot-dash blue line: \cite{bethermin2010c} model. For each model, the
  top line corresponds to all galaxies, and the bottom line to the
  prior selection (used in this work) of
  $S_{24}>80~\mu$Jy.}  \label{fig:dbdzmodel}
\end{figure*}

\subsection{The models}
\label{sect:models}

These observations can be confronted to models. We use three backwards
evolution models developed to fit infrared data: \cite{lagache2004},
\cite{le_borgne2009}, \cite{valiante2009}, and
\cite{bethermin2010c}, among an abundant literature \cite[for the most
recent:
]{franceschini2008,franceschini2009,pearson2009,rowan-robinson2009}. The
main feature of the \cite{lagache2004} model is the use of two galaxy
populations spectral energy distributions (SED), normal/cold and
starburst galaxies, parametrized by their total infrared luminosity.
The local luminosity functions are fitted, and the evolution in
redshift is applied in order to fit the additional constraints of the
observed number counts, CIB SED, and CIB fluctuation level. The
\cite{le_borgne2009} model is based on an automated
minimization of the difference between the model and selected datasets
(local luminosity functions, number counts, CIB absolute level) with a
given SED library \cite[]{chary2001}.  The \cite{valiante2009}
  model introduces scatter in the SED by using Monte-Carlo runs within
  an extended library based on observations from the Spitzer archive,
  and containing starburst and AGN contributions as a function of IR
  luminosity. Those three models used Spitzer data, and were
developed prior to the availability of Herschel data. Finally, the
\cite{bethermin2010c} model is a fully-parametric approach,
automatically fitting the latest Spitzer, BLAST \& Herschel data with
Markov chain Monte Carlo (MCMC) \cite[]{dunkley2005} method,
and using the \cite{lagache2004} SED library of two galaxy
populations.

Figures~\ref{fig:dbdz160} and \ref{fig:dbdz70} overplot these three
models in the case of a prior selection at 24~$\mu$m (with $S_{24} >
80~\mu$Jy) in order to be consistent with the data we are comparing
with. Figure~\ref{fig:dbdzmodel} shows the models, with this prior cut,
but also without any cut, i.e. all the galaxies. The main differences
can be summarized as follows:
\begin{itemize}
\item the \cite{lagache2004} model predicts a contribution of infrared
  galaxies to the 70 and 160~$\mu$m backgrounds peaking at $z \sim 1$,
  which is not observed; also the predicted dip at $z \sim 0.5$ is not
  observed either.
\item the \cite{le_borgne2009} model overpredicts the galaxy contributions
  at $z \le 0.3$, in disagreement with observations, despite the poor
  statistics of the observations; the origin is likely a lack of a
  cold galaxy component at $z \sim 0$. The general shape of the model
  at $z > 0.3$ agrees with the data, despite predicted peak and
  high-redshift ($z>1$) tail slightly higher than observed.
\item the \cite{valiante2009} model has similar trends as
    \cite{lagache2004}: almost no contribution at low redshift (in
    disagreement with data at 70 and 160~$\mu$m), and a pronounced
    peak at $z\sim 1$, not observed. This models reproduces well,
    however, the $z>1$ tail at 160~$\mu$m (but not at
    70~$\mu$m).
\item the \cite{bethermin2010c} model provides a better fit to the data,
  likely because its minimization on recent Spitzer \& Herschel data
  \cite[]{bethermin2010,oliver2010} at far-infrared and submillimeter
  wavelengths already takes into account the statistical properties of
  galaxies in an empirical way.
\item the selection of $S_{24} > 80~\mu$Jy galaxies to estimate the
  background buildup with redshift produces an almost flat cut in
  redshift to the brightness (comparison of the two lines for each
  model in fig.~\ref{fig:dbdzmodel}, except for
    \cite{valiante2009} at 160~$\mu$m with larger variations). Thus
  the peak and the structure in redshift observed with the $S_{24} >
  80~\mu$Jy cut is not much affected by this selection, and our
  conclusions based on this cut can be extended to the whole CIB
  buildup, at least for $z < 2.5$. However, the $S_{24} > 80~\mu$Jy
  cut might cause a problem of strong incompleteness at Herschel SPIRE
  wavelengths, made-up by higher-redshift sources
  \cite[]{lagache2005,marsden2009}; A need to use fainter 24~$\mu$m
  flux densities is thus required at submm wavelengths.

\end{itemize}

All these models predict similar star formation rates and luminosity
function evolutions. Our work put stronger constraints on the models,
which will have to fine-tune either the galaxies SED used or refine
the luminosity function evolutions.

\section{Conclusion}

As shown by \cite{lefloch2009} and our results, the CIB buildup allows
to break degeneracies present in the models (same predicted number
counts and CIB SED, but different redshift histories for the
luminosity functions for instance). Using exquisite Spitzer data on
one of the widest and deepest fields, we are able to measure that the
maximum contribution of the 70~$\mu$m background (as selected with
24~$\mu$m galaxies with $S_{24}>80~\mu$Jy) occurs at $z < 0.5$ and at
$0.3 < z < 0.9$ for the 160~$\mu$m background.

We measure that 74\% of the 160~$\mu$m background was emitted at $z
\le 1$, and 81\% at 70~$\mu$m. We also provided an estimate of the AGN
contribution to the far-infrared background of less than about 10\%
for $z < 1.5$.

The comparison with preliminary Herschel/PACS data on GOODS-N from
\cite{berta2010} is in line with our findings, despite the
uncertainties due to large scale structure. The consistency of the
results confirms that the stacking analysis method is a valid approach
to estimate the components of the far-IR background using prior
information on resolved mid-IR sources.

The \cite{lagache2004} model predictions mainly disagree with the
data, since the peak contribution at $z \sim 1$ is not observed. The
\cite{le_borgne2009} model disagrees with the data at low redshift
(likely due to the SED used), but succeeds in reproducing most of the
observed trend, despite an excess at $z>1.5$. The
\cite{bethermin2010c} model is favored by the data.

Our study, combined with those of \cite{lefloch2009} and
\cite{berta2010}, can allow to better constrain the models of galaxy
evolution, since their predictions can strongly vary with redshift,
despite good fits of the number counts, luminosity functions and
cosmic infrared background spectral energy distribution.

This study, together with forthcoming works to be done on Herschel
data, will also help refining the models to compute the far-infrared
and submillimeter emissivity with redshift, needed to compute the
optical depth for (hundreds of) TeV photons. Since the opacity of the
Universe for TeV photons depends on the infrared luminosity density
along the line of sight, the buildup history of the CIB has direct
effect on the TeV photons propagation. We showed that most
($\sim$80\%) of the far-infrared background is produced at $z<1$, in a
regime where many blazars are observed
\cite[e.g. ]{aharonian2006,albert2008}. The model predictions for TeV
obscuration models
\cite[e.g. ]{mazin2007,franceschini2008,stecker2009,kneiske2010,younger2010,bethermin2010c}
could be disentangled at $z \le 0.3$, where the CIB impacts high
energy photons the most, by comparing their CIB buildup history with
our data.

%
\begin{acknowledgements}
  Part of this work was supported by the D-SIGALE ANR-06-BLAN-0170 and
  the HUGE ANR-09-BLAN-0224-02.  M. J. thanks the ``Mati\`ere
  Interstellaire et Cosmologie'' group at IAS and the CNRS for the
  funding. K. C. acknowledges founding from a Leverhulme Trust Early
  Career Fellowship. We thank S. Berta for providing us with an
  electronic form of his latest Herschel/PACS PEP published results,
  and D. Le Borgne and E. Valiante for the public access to their
  models. We thank D. Elbaz, E. Daddi, B. Magnelli, and R.-R. Chary for
  fruitful discussions. {\it In memoriam, Pr. Ph. Jauzac (August 29th,
    1948 - October 22nd, 2009) (M. J.)}.
\end{acknowledgements}

\begin{table*}[!hb]
  \caption[]{The CIB brightness by redshift range at 160 $\mu$m, in units of nW.$m^{-2}$.$sr^{-1}$, in the case of the GOODS fields. The subscript ``stack'' refers to the measurement of the signal on the cleaned and stacked image; ``sources'' to the individually detected sources; ``AGN'' to the sources identified as AGN by \cite{caputi2006}. The total number of sources used in this analysis is thus $N_{stack} + N_{sources}$.}
\label{tab:CIBcontrib160GOODS}
\begin{tabular}[h]{lcccc}
\hline\\[-5pt]
 & 0$<$ $z$ $<$0.65 & 0.65$<$ $z$ $<$1.3 & 1.3$<$ $z$ $<$2 & $z$ $>$2 \\
\hline\\[-5pt]
$N_{stack}$ & 317 & 573 & 258 & 198 \\ 
$N_{sources}$ & 10 & 11 & 3 & 2 \\
$N_{AGNs}$ & 9 & 45 & 38 & 44 \\
\hline\\[-5pt]
$B_{160_{stack}}$ & 2.18 $\pm$ 0.45 & 2.39 $\pm$ 0.54 & 1.13 $\pm$ 0.35 & 0.78 $\pm$ 0.27 \\
$B_{160_{sources}}$ & 0.57 $\pm$ 0.18 & 0.34 $\pm$ 0.11 & 0.06 $\pm$ 0.04 & 0.07 $\pm$ 0.05 \\
$B_{160_{AGNs}}$ & 0.04 $\pm$ 0.07 & 0.17 $\pm$ 0.25 & 0.32 $\pm$ 0.22 & 0.20 $\pm$ 0.19 \\
$B_{160_{tot}}$ & 2.75 $\pm$ 0.46 & 2.73 $\pm$ 0.54 & 1.19 $\pm$ 0.35 & 0.85 $\pm$ 0.27 \\
\hline\\[-5pt]
\end{tabular}
\end{table*}

\begin{table*}[!hb]
  \caption[]{The CIB brightness by redshift range at 70 $\mu$m, in units of nW.$m^{-2}$.$sr^{-1}$, in the  GOODS fields. Terms are defined in the caption of Tab.~\ref{tab:CIBcontrib160GOODS}.}
\label{tab:CIBcontrib070GOODS}
\begin{tabular}[h]{lcccc}
\hline\\[-5pt]
 & 0$<$ $z$ $<$0.65 & 0.65$<$ $z$ $<$1.3 & 1.3$<$ $z$ $<$2 & $z$ $>$2 \\
\hline\\[-5pt]
$N_{stack}$ & 317 & 575 & 259 & 198 \\ 
$N_{sources}$ & 19 & 5 & 1 & 0 \\
$N_{AGNs}$ & 9 & 45 & 38 & 44 \\
\hline\\[-5pt]
$B_{70_{stack}}$ & 1.29 $\pm$ 0.22 & 1.25 $\pm$ 0.28 & 0.35 $\pm$ 0.15 & 0.15 $\pm$ 0.12 \\
$B_{70_{sources}}$ & 0.77 $\pm$ 0.18 & 0.14 $\pm$ 0.07 & 0.01 $\pm$ 0.015 & -- \\
$B_{70_{AGNs}}$ & 0.05 $\pm$ 0.03 & 0.1 $\pm$ 0.12 & 0.11 $\pm$ 0.08 & 0.11 $\pm$ 0.12 \\
$B_{70_{tot}}$ & 2.06 $\pm$ 0.24 & 1.4 $\pm$ 0.28 & 0.36 $\pm$ 0.15 & 0.15 $\pm$ 0.12 \\
\hline\\[-5pt]
\end{tabular}
\end{table*}

\begin{table*}[!hb]
  \caption[]{The CIB brightness by redshift range at 160 $\mu$m, in units of nW.$m^{-2}$.$sr^{-1}$, in the case of the COSMOS field. The subscript ``stack'' refers to the measurement of the signal on the cleaned and stacked image; ``sources'' to the individually detected sources; ``AGN'' to the sources identified as AGN by \cite{salvato2009}. The total number of sources used in this analysis is thus $N_{stack} + N_{sources}$.}
\label{tab:CIBcontrib160COSMOS}
\begin{tabular}[h]{lccccccc}
\hline\\[-5pt]
 & 0$<$ $z$ $<$0.15 & 0.15$<$ $z$ $<$0.3 & 0.3$<$ $z$ $<$0.45 & 0.45$<$ $z$ $<$0.6 & 0.6$<$ $z$ $<$0.75 & 0.75$<$ $z$ $<$0.9 & 0.9$<$ $z$ $<$1.05 \\
\hline\\[-5pt]
$N_{stack}$ & 2083 & 1559 & 2853 & 2201 & 3225 & 3590 & 3478 \\ 
$N_{sources}$ & 56 & 40 & 18 & 9 & 11 & 12 & 6 \\
$N_{AGNs}$ & 34 & 32 & 74 & 48 & 88 & 109 & 123 \\
\hline\\[-5pt]
$B_{160_{stack}}$ & 0.43 $\pm$ 0.04 & 0.56 $\pm$ 0.04 & 0.87 $\pm$ 0.05 & 0.63 $\pm$ 0.04 & 0.76 $\pm$ 0.07 & 0.75 $\pm$ 0.06 & 0.71 $\pm$ 0.06 \\
$B_{160_{sources}}$ & 0.3 $\pm$ 0.001 & 0.21 $\pm$ 0.002 & 0.07 $\pm$ 0.001 & 0.03 $\pm$ 0.001 & 0.034 $\pm$ 0.001 & 0.043 $\pm$ 0.001 & 0.021 $\pm$ 0.001 \\
$B_{160_{AGNs}}$ & 0.02 $\pm$ 0.006 & 0.01 $\pm$ 0.005 & 0.03 $\pm$ 0.01 & 0.02 $\pm$ 0.01 & 0.02 $\pm$ 0.01 & 0.01 $\pm$ 0.01 & 0.02 $\pm$ 0.01 \\
$B_{160_{tot}}$ & 0.72 $\pm$ 0.04 & 0.77 $\pm$ 0.04 & 0.93 $\pm$ 0.05 & 0.66 $\pm$ 0.04 & 0.8 $\pm$ 0.07 & 0.8 $\pm$ 0.06 & 0.73 $\pm$ 0.06 \\
\hline\\[-5pt]
\hline\\[-5pt]
 & 1.05$<$ $z$ $<$1.2 & 1.2$<$ $z$ $<$ 1.35 & 1.35$<$ $z$ $<$1.5 & 1.5$<$ $z$ $<$1.65 & 1.65$<$ $z$ $<$1.85 & 1.85$<$ $z$ $<$2.05 & $z$ $>$2.05 \\
 \hline\\[-5pt]
 $N_{stack}$ & 2670 & 1401 & 2044 & 1311 & 1519 & 2073 & 2833 \\ 
$N_{sources}$ & 2 & 0 & 2 & 2 & 0 & 1 & 2 \\
$N_{AGNs}$ & 83 & 76 & 55 & 225 & 230 & 198 & 288 \\
\hline\\[-5pt]
$B_{160_{stack}}$ & 0.52 $\pm$ 0.05 & 0.33 $\pm$ 0.09 & 0.36 $\pm$ 0.04 & 0.26 $\pm$ 0.03 & 0.22 $\pm$ 0.02 & 0.27 $\pm$ 0.04 & 0.47 $\pm$ 0.04 \\
$B_{160_{sources}}$ & 0.005 $\pm$ 0.0003 & -- & 0.006 $\pm$ 0.0003 & 0.006 $\pm$ 0.0003 & -- & 0.003 $\pm$ 0.0001 & 0.007 $\pm$ 0.0003 \\
$B_{160_{AGNs}}$ & 0.02 $\pm$ 0.01 & 0.02 $\pm$ 0.01 & 0.01 $\pm$ 0.008 & 0.04 $\pm$ 0.03 & 0.04 $\pm$ 0.03 & 0.05 $\pm$ 0.03 & 0.04 $\pm$ 0.03  \\
$B_{160_{tot}}$ & 0.53 $\pm$ 0.05 & 0.33 $\pm$ 0.09 & 0.37 $\pm$ 0.04 & 0.27 $\pm$ 0.03 & 0.22 $\pm$ 0.02 & 0.28 $\pm$ 0.04 & 0.47 $\pm$ 0.04 \\
\hline\\[-5pt]
\end{tabular}
\end{table*}

\begin{table*}[!hb]
  \caption[]{The CIB brightness by redshift range at 70 $\mu$m, in units of nW.$m^{-2}$.$sr^{-1}$ , in the COSMOS field. Terms are defined in the caption of Tab.~\ref{tab:CIBcontrib160COSMOS}.}
\label{tab:CIBcontrib070COSMOS}
\begin{tabular}[h]{lccccccc}
\hline\\[-5pt]
 & 0$<$ $z$ $<$0.15 & 0.15$<$ $z$ $<$0.3 & 0.3$<$ $z$ $<$0.45 & 0.45$<$ $z$ $<$0.6 & 0.6$<$ $z$ $<$0.75 & 0.75$<$ $z$ $<$0.9 & 0.9$<$ $z$ $<$1.05 \\
\hline\\[-5pt]
$N_{stack}$  & 2083 & 1559 & 2853 & 2202 & 3225 & 3590 & 3478 \\ 
$N_{sources}$ & 77  & 82   & 48   & 23   & 13   & 11   & 3 \\
$N_{AGNs}$   & 34   & 32   & 74   & 48   & 88  & 110   & 123 \\
\hline\\[-5pt]
$B_{70_{stack}}$ & 0.31 $\pm$ 0.03 & 0.39 $\pm$ 0.02 & 0.6 $\pm$ 0.03 & 0.42 $\pm$ 0.03 & 0.47 $\pm$ 0.03 & 0.41 $\pm$ 0.03 & 0.39 $\pm$ 0.03 \\
$B_{70_{sources}}$ & 0.31 $\pm$ 0.001 & 0.34 $\pm$ 0.001 & 0.13 $\pm$ 0.0005 & 0.05 $\pm$ 0.0003 & 0.04 $\pm$ 0.0003 & 0.027 $\pm$ 0.0003 & 0.008 $\pm$ 0.0001 \\
$B_{70_{AGNs}}$ & 0.011 $\pm$ 0.004 & 0.008 $\pm$ 0.003 & 0.015 $\pm$ 0.006 & 0.009 $\pm$ 0.003 & 0.016 $\pm$ 0.005 & 0.015 $\pm$ 0.007 & 0.015 $\pm$ 0.008 \\
$B_{70_{tot}}$ & 0.62 $\pm$ 0.03 & 0.73 $\pm$ 0.02 & 0.73 $\pm$ 0.03 & 0.47 $\pm$ 0.03 & 0.51 $\pm$ 0.03 & 0.44 $\pm$ 0.03 & 0.4 $\pm$ 0.03 \\
\hline\\[-5pt]
\hline\\[-5pt]
 & 1.05$<$ $z$ $<$1.2 & 1.2$<$ $z$ $<$ 1.35 & 1.35$<$ $z$ $<$1.5 & 1.5$<$ $z$ $<$1.65 & 1.65$<$ $z$ $<$1.85 & 1.85$<$ $z$ $<$2.05 & $z$ $>$2.05 \\
\hline\\[-5pt]
$N_{stack}$ & 2670 & 1401 & 2044 & 1311 & 1519 & 2073 & 2833 \\ 
$N_{sources}$ & 2    & 1    & 2    & 0    & 0    & 0    & 0 \\
$N_{AGNs}$ & 83   & 76   & 55   & 225  & 232  & 200  & 288 \\
\hline\\[-5pt]
$B_{70_{stack}}$ & 0.25 $\pm$ 0.02 & 0.18 $\pm$ 0.01 & 0.17 $\pm$ 0.02 & 0.1 $\pm$ 0.01 & 0.09 $\pm$ 0.02 & 0.09 $\pm$ 0.02 & 0.16 $\pm$ 0.02 \\
$B_{70_{sources}}$ & 0.004 $\pm$ 0.0001 & 0.002 $\pm$  0.0001 & 0.003 $\pm$ 0.0001 & -- & -- & -- & -- \\
$B_{70_{AGNs}}$ & 0.14 $\pm$ 0.007 & 0.011 $\pm$ 0.005 & 0.006 $\pm$ 0.004 & 0.28 $\pm$ 0.012 & 0.026 $\pm$ 0.015 & 0.022 $\pm$ 0.01 & 0.029 $\pm$ 0.018 \\
$B_{70_{tot}}$ & 0.26 $\pm$ 0.02 & 0.18 $\pm$ 0.01 & 0.17 $\pm$ 0.02 & 0.1 $\pm$ 0.01 & 0.09 $\pm$ 0.2 & 0.09 $\pm$ 0.02 & 0.15 $\pm$ 0.02 \\
\hline\\[-5pt]
\end{tabular}
\end{table*}


\begin{table*}[b]
  \caption[]{The total CIB brightness at 160 $\mu$m \& 70$\mu$m for the GOODS \& COSMOS fields, in units of nW.$m^{-2}$.$sr^{-1}$:
 lines 1 \& 2: our estimates; 
line 3: \cite{bethermin2010} CIB measured value using number counts integration;
line 4: \cite{bethermin2010} CIB value with extrapolation of the number counts in power-law;
line 5: \cite{lagache2004} CIB model value with the constraint: $S_{24}$ $>$ 80 $\mu$Jy;
line 6: \cite{lagache2004} CIB model value of the total background; 
line 7: \cite{le_borgne2009} CIB model value with the constraint: $S_{24}$ $>$ 80 $\mu$Jy; 
line 8: \cite{le_borgne2009} CIB model value of the total background; 
line 9: \cite{valiante2009} CIB model value with the constraint: $S_{24}$ $>$ 80 $\mu$Jy; 
line 10: \cite{valiante2009} CIB model value of the total background; 
line 11: \cite{bethermin2010c} CIB model value with the constraint: $S_{24}$ $>$ 80 $\mu$Jy; 
line 12: \cite{bethermin2010c} CIB model value of the total background.
}
\label{tab:CIBtot}
\begin{center}
\begin{tabular}[h]{lccccc}
\hline\\[-5pt]
 & 160 $\mu$m & 70 $\mu$m \\
\hline\\[-5pt]
$B_{tot-GOODS}$ & 7.53 $\pm$ 0.84 & 3.97 $\pm$ 0.41 \\
$B_{tot-COSMOS}$ & 7.88 $\pm$ 0.19 & 4.95 $\pm$ 0.08 \\
$B_{Bethermin}$ & 9.0 $\pm$ 1.1 & 5.2 $\pm$ 0.4 \\
$B_{Bethermin_{CIB estimate}}$ & $14.6^{+7.1}_{-2.9}$ & $6.4^{+0.7}_{-0.6}$ \\
$B_{model_{Lagache}}(S_{24}>80 \mu Jy)$ & 11.91 & 5.73 \\
$B_{model_{Lagache}} $ & 14.87 & 6.78 \\
$B_{model_{LeBorgne}}(S_{24}>80 \mu Jy)$ & 9.54 & 6.65 \\
$B_{model_{LeBorgne}}$ & 13.57 & 8.54 \\
$B_{model_{Valiante}}(S_{24}>80 \mu Jy)$ & 6.84 & 4.27 \\
$B_{model_{Valiante}}$ & 16.70 & 6.98 \\
$B_{model_{Bethermin}}(S_{24}>80 \mu Jy)$ & 8.82  & 5.02\\ 
$B_{model_{Bethermin}}$ & 11.66  & 6.09 \\

\hline\\[-5pt]
\end{tabular}
\end{center}
\end{table*}

%


\end{document}